\documentclass[12pt,letterpaper]{JHEP3}
\usepackage{amsmath,amssymb,amsfonts,xspace}

\def\Re{\,{\rm Re}\, }

\def\({\left(}
\def\){\right)}
\def\[{\left[}
\def\]{\right]}

\newcommand{\de}{\mathrm{d}}

\newcommand{\I}{\mathrm{i}}

\newcommand{\cF}{\mathcal{F}}

\newcommand{\cC}{\mathcal{C}}

\newcommand{\cM}{\mathcal{M}}

\newcommand{\cR}{\mathcal{R}}

\DeclareSymbolFont{AMSa}{U}{msa}{m}{n}
\DeclareSymbolFont{AMSb}{U}{msb}{m}{n}
\DeclareMathSymbol{\fieldR}{\mathalpha}{AMSb}{"52}

\newcommand{\cO}{\mathcal{O}}

\renewcommand{\Re}{{\rm Re}}

\newcommand{\pa}{\partial}

\newcommand{\IR}{\mathbb{R}}
\newcommand{\IC}{\mathbb{C}}
\newcommand{\IZ}{\mathbb{Z}}

\newcommand{\IQ}{\mathbb{Q}}
\newcommand{\Zint}{\mathbb{Z}}

\def\bea{\begin{eqnarray}}
\def\eea{\end{eqnarray}}
\def\be{\begin{equation}}
\def\ee{\end{equation}}
\def\ba{\begin{align}}
\def\ea{\end{align}}
\def\bse{\begin{subequations}}
\def\ese{\end{subequations}}

\newcommand{\irrep}[1]{\ensuremath{\boldsymbol{#1}}}
\newcommand{\eis}[3]{~\ensuremath{{\cal E}^{#1}_{\irrep{#2};#3}}}

\newcommand{\vect}{\ensuremath{{\bf V}}}
\newcommand{\spi}{\ensuremath{{\bf S}}}
\newcommand{\spb}{\ensuremath{{\bf C}}}

\newcommand{\eishat}[3]{~\ensuremath{\hat {\cal E}^{#1}_{\irrep{#2};#3}}}
\newcommand{\eisstar}[3]{~\ensuremath{{\cal E}^{#1,\star}_{\irrep{#2};#3}}}
\newcommand{\eishatstar}[3]{~\ensuremath{\hat {\cal E}^{#1,\star}_{\irrep{#2};#3}}}

\DeclareMathOperator{\Ind}{Ind} 
\DeclareMathOperator{\Res}{Res} 
\newcommand{\zetastar}{\zeta^\star}
\newcommand{\url}[1]{{\tt#1 } \href{#1}}

\title{$R^4$ couplings and automorphic unipotent representations}

\author{Boris Pioline
\\
{\it Laboratoire de Physique Th\'eorique et Hautes
Energies, \\ CNRS UMR 7589 and  
Universit\'e Pierre et Marie Curie - Paris 6,\\
4 place Jussieu, 75252 Paris cedex 05, France} \\
Email: \email{pioline@lpthe.jussieu.fr}
           }

\abstract{Four-graviton, eight-derivative couplings in the low energy effective action
of toroidal type II string compactifications are tightly constrained 
by U-duality invariance 
and by supersymmetry. In this note, we revisit earlier proposals for the automorphic form 
governing these couplings in dimension $D=3,4,5,6$, and propose that 
the correct automorphic form is the minimal theta series for the corresponding
U-duality group. Evidence for this proposal comes from i) the matching of
infinitesimal characters, ii) the fact that the Fourier coefficients have support
on 1/2-BPS charges and iii) decompactification limits. In particular,
we show that non-perturbative effects can be interpreted as 1/2-BPS instantons, or 1/2-BPS 
particles in one dimension higher (together with Taub-NUT instantons in the $D=3$ case).
Based on similar considerations, we also conjecture the form of 1/4-BPS saturated couplings
such as $\nabla^4 R^4$ couplings in the same dimensions.
}


\preprint{arXiv:1001.3647v3}

\begin{document}

\section{Introduction}

In the absence of a non-perturbative formulation of string theory, 
U-dualities \cite{Hull:1994ys}
offer a Unique tool for computing exact amplitudes, including instanton effects. This programme has been particularly successful in the case of four-graviton couplings in the low energy effective action 
around string vacua with maximal supersymmetry. 

For type IIB string theory in 10 
dimensions, an exact answer for the eight-derivative, $R^4$ coupling consistent with duality invariance 
under $G_{10}=SL(2,\IZ)$ was proposed in the seminal work \cite{Green:1997tv}, and verified in a number of subsequent works \cite{Green:1997as,Pioline:1998mn,Green:1998by,Moore:1998et}.
This was extended to toroidal compactifications of M-theory on $T^2$ \cite{Green:2005ba}, 
$T^3$ \cite{Green:2005ba, Kiritsis:1997em} and $T^4$ \cite{Kiritsis:1997em}, consistently
with the  U-duality symmetries $G_9=SL(2,\IZ)$, $G_8=SL(2,\IZ)\times SL(3,\IZ)$ and 
$G_7=SL(5,\IZ)$, respectively. In all these cases, the $R^4$ coupling $f_{R^4}^{(D)}$ 
is given by an Epstein
Zeta series (a special kind of Eisenstein series) of order $s=3/2$ for the relevant U-duality 
group $G_D$, and is an eigenmode of the Laplace-Beltrami operator on the moduli space $K_D\backslash G_D$, as required by supersymmetry \cite{Pioline:1998mn,Green:1998by}.

For toroidal compactifications to lower 
dimension $3\leq D\leq 6$, it was pointed out in \cite{Obers:1999um} that  
the Epstein zeta series of  $G_D=E_{11-D}$ could no longer be the right object to represent
$f_{R^4}^{(D)}$, as they are not eigenmodes of the Laplacian;  it was proposed to remedy
this deficiency by considering constrained Epstein zeta series built out of certain finite-dimensional
representations $\cR$ of $G_D$. A selection of pairs $(\cR,s)$ was proposed to represent
$f_{R^4}^{(D)}$ in all dimensions $D\geq 3$, as summarized in Table \ref{table1}.
 
In this note, we revisit the claim in \cite{Obers:1999um},
and argue that the appropriate automorphic form describing the $R^4$
couplings in dimension $3\leq D\leq 6$ is in fact the (non-Gaussian) theta 
series $\theta_{G_D}$ associated to the minimal unitary representation of $G_D$.
This proposal is based on identifying the infinitesimal character  (which encodes
the eigenvalues of $f_{R^4}^{(D)}$ under all $G_D$-invariant differential operators), and on 
unique properties of the Fourier coefficients of the minimal theta series. In particular, we
show that all non-perturbative contributions to $\theta_{G_D}$ can be interpreted as 
coming from 1/2-BPS instantons, or from 1/2-BPS particles in $D+1$ dimensions. 
Using similar arguments, we also suggest that the 12-derivative $\nabla^4 R^4$
coupling in dimension $D=3,4$ are governed by the``next-to-minimal" unipotent
representation constructed in \cite{MR1421947,MR1327538}, whose Fourier
coefficients have support on 1/4-BPS charges. Unfortunately, we do not have
a concrete automorphic form at hand in this case.

We note that the relevance of unipotent representations for black hole
partition functions and instanton sums has been suggested in earlier 
works \cite{Pioline:2001jn,Pioline:2004xq,Pioline:2005vi,Gunaydin:2005mx, Pioline:2009qt} 
(see also \cite{Gunaydin:2000xr,Gunaydin:2001bt}). The present work improves
on these earlier attempts by identifying concrete couplings in the low energy effective
action described by these automorphic representations.

This note is organized as follows. In Section \ref{secrev} we review the proposal in \cite{Obers:1999um},
and determine the infinitesimal character $\rho+\lambda$  associated to the $R^4$ couplings in dimension $3\leq D\leq 6$. In Section \ref{secr4} we formulate our conjecture, and analyze
its implications for the form of the non-perturbative contributions to $f_{R^4}^{(D)}$ in various
limits. In Section 
\ref{sec_d4r4} we discuss the case of $\nabla^4 R^4$ couplings and conclude.

While the first version of this note was being written, two papers appeared which 
had some overlap with the present work \cite{Green:2010wi,LambertWest}.
The subsequent release of \cite{Green:2010sp} prompted me to further
study the consistency of the conjectures herein under decompactification from $D$
to $D+1$ dimensions. The results of this investigation, based on  \cite{Green:2010wi,Green:2010sp} and results available in the mathematical literature \cite{MR1159110,MR1469105},
are presented in an Appendix.

\begin{table}
$$
\begin{array}{c@{\hspace{4mm}}c@{\hspace{4mm}}lr@{\hspace{4mm}}c@{\hspace{4mm}}l}
\hline
D & G_D &\cR  & {\rm dim}\, \cR & s & \rho+\lambda \\
   \hline
 6 & SO(5,5) &   \mbox{string}   & \irrep{10} & 3/2 & [-2,1,1,1,1]\\
   & & \mbox{particle}   & \irrep{16} & 1 & [1,1,1,1,-1]\\
   &      & \mbox{membrane}&  \irrep{16'} &1 & [1,1,1,-1,1]  \\
        &          & \mbox{5-brane}&  \irrep{120} & 2 &  [1,1,0,1,1]  \\
\hline
5 & E_{6(6)}  &\mbox{string}   &  \irrep{27} & 3/2 &[-2,1,1,1,1,1]\\
   & & \mbox{particle}   & \irrep{27'} & 3/2 & [1,1,1,1,1,-2] \\
   & & \mbox{membrane}&   \irrep{78} & 1 & [1,-1,1,1,1,1] \\
      &          & \mbox{5-brane}&  \irrep{25.920} &2 & [1,1,1,0,1,1]  \\
   \hline
4  & E_{7(7)} &     \mbox{string}   &  \irrep{133} & 3/2  & [-2,1,1,1,1,1,1]\\
  &          & \mbox{particle}   &  \irrep{56} & 2 & [1,1,1,1,1,1,-3]\\
  &      & \mbox{membrane}&  \irrep{912} & 1 & [1,-1,1,1,1,1,1]  \\
     &          & \mbox{5-brane}&  \irrep{365.750} & 2 & [1,1,1,0,1,1,1]  \\
  \hline
 3 & E_{8(8)} & \mbox{string}   &  \irrep{3875} & 3/2 & [-2,1,1,1,1,1,1,1]\\
   &          & \mbox{particle}   &  \irrep{248} & 5/2 & [1,1,1,1,1,1,1,-4]\\
   &          & \mbox{membrane}&  \irrep{147.250} & 1 &[1,-1,1,1,1,1,1,1] \\
   &          & \mbox{5-brane}&  \irrep{6.899.079.264}    & 2 &[1,1,1,0,1,1,1,1]  \\
\hline
\end{array}
$$
\caption{Constrained Epstein zeta series proposed to describe $f_{R^4}^{(D)}$. Entries
corresponding to the``5-brane" multiplet (corresponding to the fundamental weight
associated to the trivalent node in the Dynkin diagram, according to the terminology of
\cite{Obers:1998fb}) were not considered in \cite{Obers:1999um}, but enter the discussion
in Section 3. The weight vector $\rho+\lambda$ is described by its coordinates in fundamental weight basis, using the same conventions as in LiE.\label{table1}}
\end{table}

\section{Constrained zeta series and infinitesimal characters
\label{secrev}}

Let us start by reviewing the main argument of \cite{Obers:1999um}. From the tree-level
and one-loop computations in \cite{Gross:1986iv} and \cite{Kiritsis:1997em}, it is known that
the $R^4$ coupling in type II string theory compactified on a torus $T^d$ must be given, 
at weak 10D string coupling $g_s$, by
\be
\label{fr4pert}
f_{R^4}^{(D)} = 2\zeta(3) \frac{V_d}{g_s^2 l_s^2} + \frac{I_d}{l_s^2} + \cO(g_s^2)
\ee
where $I_d$ is the integral of the partition function of the Narain lattice of signature $(d,d)$ 
on the fundamental domain $\cF$ of the Poincar\'e upper half plane,
\be
\label{defId}
I_d = 2\pi \int_\cF\, \frac{\de\tau_1\de\tau_2}{\tau_2^2}\, (Z_{d,d}(g,B;\tau) -\tau_2^{d/2} )\ .
\ee
The subtraction of the second term (omitted in \cite{Obers:1999um}) renders the integral
convergent in the infrared region $\tau_2\to\infty$, at the cost of breaking modular invariance. 
Other regulators would differ from $I_d$ by an  additive, moduli independent constant which
we ignore.
It was observed in \cite{Obers:1999um} that the one-loop contribution $I_d$ could be written
as
\be
\label{conjv}
I_d = \frac{2\, \Gamma\left(\frac{d}{2}-1\right)}{\pi^{\frac{d}{2}-2}}\,  \eis{SO(d,d,\IZ)}{\vect}{s=\frac{d}{2}-1}\ ,
\ee
where \eis{SO(d,d,\IZ)}{\vect}{s} is 
the  constrained Epstein zeta series for the group $SO(d,d,\IR)$, 
built out of the $2d$-dimensional  (vector representation $\cR=\vect$,
\be
\label{defeisv}
\eis{SO(d,d,\IZ)}{\vect}{s} = \sum_{\substack{ m_i\in\IZ^d, n^i \in\IZ^d\\ (m_i,n^i)\neq (0,0);
m_i n^i=0}}
\left[ \left(m_i + B_{ik} n^k\right) g^{ij} \left(m_j + B_{jl} n^l\right) + n^i g_{ij} n^j \right]^{-2s} \ .
\ee
This identity was suggested by comparing the eigenvalues under the Laplacian on 
$[SO(d)\times SO(d)]\backslash SO(d,d,\IR)$, as well as under a non-invariant operator 
$\square_d$, and established by analyzing the large volume behavior of both sides,
\eqref{defeisv} being defined away from its domain of absolute convergence 
by analytic continuation in $s$. Similar identities were proposed involving the 
constrained Epstein zeta series constructed out of the spinor representations $\spi$
and $\spb$, 
\begin{equation}
\label{conj1}
I_d = 2 \eis{SO(d,d,\Zint)}{\spi}{s=1} = 2 \eis{SO(d,d,\Zint)}{\spb}{s=1}\ ,\qquad (d>2)
\end{equation}
but were left as conjectures due to the difficulty in analyzing the large volume
asymptotics of these constrained Epstein zeta series.

According to the U-duality hypothesis, the non-perturbative completion of the $R^4$ coupling  \eqref{fr4pert}  should be invariant under a larger group $G_{10-d}$, generated by two non-commuting actions, T-duality $SO(d,d,\IZ)$ and M-theory diffeomorphism invariance $SL(d+1,\IZ)$  \cite{Hull:1994ys,Obers:1998fb}. Since the first (second) term of \eqref{fr4pert} can itself be written as an Epstein zeta series in the singlet (spinor $\spi$) representation, one (admittedly simple-minded)
strategy is to find a representation $\cR$ of $G_{10-d}$ which decomposes as $1+ \spi+\dots$ under $SO(d,d,\IR)$. 
By following this route, we
were led in  \cite{Obers:1999um} to propose that the exact coupling $f_{R^4}^{(D)}$ should be given, up 
to a numerical factor and power of the U-duality invariant Planck scale 
$m_p = (V_{d}/g_s^2 l_s^8)^{1/(8-d)}$, by a constrained Epstein zeta series of $G_{10-d}$
with representation $\cR$ of highest weight $\lambda_\cR$ and order $s\in \IC$, defined as 
\be
\label{defceis}
\eis{G_{D}}{\cR}{s} = \sum_{m\in \cC-\{0\}} \left[ m^t \cdot \cM \cdot m\right]^{-s}\ .
\ee
Here the sum runs over non-zero integer vectors $m\in \Lambda$ in a lattice $\Lambda\subset 
\IZ^{{\rm dim} \cR}$ invariant 
under $G_D(\IZ)$, and is restricted to a cone $\cC\subset \IR^{{\rm dim} \cR}$ 
such that
all symmetric powers $\vee^N m$ lie entirely in the representation of highest weight $N\lambda_\cR$, for all $N>1$ and $m\in\IC$. The series \eqref{defceis} converges absolutely when $\Re(s)$ is large
enough (in particular when $\Re(s)>{\rm dim} \cR/2$),  is by construction invariant under $G_D(\IZ)$, 
and is an eigenmode of all
invariant differential operators. In particular, the Epstein series in Table \ref{table1}
are eigenmodes of the Laplacian operator on $K_D\backslash G_D$, with eigenvalue
\begin{equation}
\label{npeig}
\Delta_{K_D\backslash G_D} f_{R^4}^{(D)} = \frac{3(d+1)(2-d)}{2(8-d)} f_{R^4}^{(D)}\ .
\end{equation}
This property holds order by order  in the asymptotic expansion at weak coupling, and 
in particular for the tree-level term in \eqref{fr4pert}.

The values of $(\cR,s)$ in Table \ref{table1} are not in the domain of absolute convergence of the 
constrained Epstein series. With the exception of the constrained Epstein series for $SO(5,5)$
in the vector representation, to be discussed below, we have at present little control on the 
analytic structure of these series, and it is therefore difficult to test these 
conjectures.
However, one useful 
fact that can be drawn from them is the infinitesimal character $\rho+\lambda$ associated 
to $f_{R^4}^{(D)}$: recall that according to the Harish-Chandra homomorphism, the  orbit of
the infinitesimal character $\rho+\lambda$ of an irreducible representation $\cR$  under the Weyl group 
encodes the value of all the Casimir operators, in particular the quadratic Casimir 
\be
\label{c2rho}
C_2  =  (\rho+\lambda,\rho+\lambda)-(\rho,\rho)\ ,
\ee
which is manifestly Weyl-invariant\footnote{Recall that the Weyl group acts by orthogonal reflections
on $\rho+\lambda$, leaving the Killing norm $(\rho+\lambda,\rho+\lambda)$ invariant.}. 
To each of these Casimir operators corresponds an  invariant differential operator 
on $K_D\backslash G_D$ (in our normalization, $\Delta=C_2/4$). 
In the case at hand, the same technics as in  \cite{Obers:1999um} show that the constrained Epstein zeta series built from the finite-dimensional 
representation is an eigenvalue of all the invariant differential operators, with infinitesimal
character 
\be
\rho+\lambda=\rho-2 s \lambda_\cR\ .
\ee
This tabulated in the last 
column of Table \ref{table1} for all Epstein zeta series relevant for $R^4$ couplings\footnote{Recall 
that the Weyl vector is the sum of all fundamental weights, $\rho=[1,1,\dots,1]$. For $\cR=\irrep{10}$
of $SO(5,5)$, $\lambda_\cR=[1,0,0,0,0]$ hence $\rho+\lambda=[-2,1,1,1,1]$, etc.}.
As a consistency check, one may verify that \eqref{c2rho}
reproduces (four times) the eigenvalue in \eqref{npeig} for all entries in the table. Moreover,
for each value of $D$, the infinitesimal characters of the four alternative representations turn
out to be in the same Weyl orbit, as must be the case if these alternatives are to describe the 
same automorphic form $f_{R^4}^{(D)}$. The  infinitesimal character associated 
to the ``5-brane multiplet" (not discussed in \cite{Obers:1999um})
turns out to be the dominant weight in this Weyl orbit, i.e. the only one with non-negative coefficients
in the fundamental weight basis.

\section{$R^4$ couplings  and minimal representations \label{secr4}}

The first point of this note is to observe that the infinitesimal characters appearing in 
Table \ref{table1} are in fact the ones associated to the minimal representation of $G_D$.
Indeed,  the infinitesimal
character in the  ``5-brane multiplet" representation, in the same Weyl orbit as the string, particle and 
membrane representations, precisely matches the result in Proposition 6.4 in \cite{MR0404366}.
For these very special values of the Casimirs, it becomes possible for the 
eigenform $f_{R^4}^{(D)}$ to be annihilated by a much larger ideal in the universal
enveloping algebra of $G_D$ than the one given by the Harish-Chandra homomorphism, 
namely Joseph's ideal \cite{MR0404366} (in which case 
the automorphic form is said  to be associated to 
the minimal representation of $G$). While generic Eisenstein series are only annihilated
by their Harish-Chandra ideal, residues of Eisenstein series 
typically exhibit an enlarged annihilator, and the minimal representation arises in a sense
as the most singular residue, with the largest annihilator \cite{MR1469105}. 

This enhancement is in fact known to take place for the  constrained Epstein zeta series of
$SO(5,5)$ (i.e. $D=6$) in the vector representation 10 at $s=3/2$: applying 
\eqref{conjv} for $d=5$, we see that this constrained Epstein zeta is the same as the 
integral of the partition function of a  (non-perturbative) Narain lattice $\Gamma_{5,5}$
over the fundamental domain. As explained in \cite{Kazhdan:2001nx} (and
well-known to mathematicians), for any $d$ 
the integral $I_d$ is in fact a (non-Gaussian) theta series associated to the 
minimal representation of $SO(d,d,\IR)$. In particular, the  worldsheet instanton sum
is a sum over rank 2 antisymmetric $n\times n$ matrices $m^{ij}$, a cone of dimension 
$2d-3$ which is the dimension of the minimal representation of $SO(d,d)$.

This enhancement is also  desirable for the following physical reason. By the standard
argument based on counting fermionic zero modes (see e.g. \cite{Kiritsis:1999ss}), we expect
that instanton corrections to eight-derivative couplings around a vacuum with 32 unbroken supercharges should originate from 1/2-BPS instanton configurations only. This implies that the
Fourier expansion of $f_{R^4}^{(D)}$ with respect to axionic scalars should have support on
a restricted set of charges. It is often the case that this condition can be expressed by
a set of differential operators annihilating the automorphic function of interest. Each
of these differential equations is not necessarily invariant under $G_D$, but their intersection 
defines an ideal in the universal enveloping algebra of $G_D$ which is invariant under the
left and
right actions of $G_D$. Joseph's ideal is the largest such ideal, and is expected to 
control 1/2-BPS saturated couplings such as $f_{R^4}^{(D)}$.

Thus, it is natural to conjecture that the exact $R^4$ coupling in dimension $D$ are in fact given 
by the (non-Gaussian) theta series $\theta_G$ associated to the minimal representation 
of $G_D$\footnote{It is likely that this theta series be identical to the constrained Epstein zeta series appearing in Table \ref{table1} after analytic  continuation in the $s$-plane, but we do not now how to prove this
except for $D=6$.},
\be
f_{R^4}^{(D)} \propto \theta_{G_D}\ . \label{conj}
\ee
To check this proposal and fix the normalization, 
one should compute the weak coupling expansion of $\theta_G$ and
 check that it reproduces the correct perturbative terms in \eqref{fr4pert} 
 (while the infinitesimal character essentially ensures the right functional dependence of these terms, 
it would be a very non-trivial check  to match the numerical coefficients). This can be easily been checked for $D=6$ using techniques in \cite{Obers:1999um}, but we are not yet able to carry out this
computation for $D=3,4,5$. In this rest of this note, we shall instead focus on the non-perturbative
effects predicted by our proposal.

\section{Analysis of non-perturbative effects \label{sec_np}}

Relying on results obtained in \cite{Kazhdan:2001nx} (see also 
\cite{MR2290764}), we now discuss the form of the non-perturbative
contributions predicted by our proposal \eqref{conj}.
We consider three different limits: 
\begin{itemize}
\item[i)] 
the decompactification limit, where one of the circles of $T^{d+1}$ becomes
large; 
\item[ii)] the string theory limit, where the  $D$-dimensional 
string coupling $g_D = g_s / \sqrt{V_d/l_s^d}$ goes to zero (we consider only
the type IIB description); 
\item[iii)]
 the M-theory limit, where the volume of the torus $T^{d+1}$ in 11D Planck units $V_{d+1}/l_M^{d+1}$ 
becomes large. 
\end{itemize}
These limits  are associated to parabolic subgroups $P$ of 
$G_D$ with Levi subgroup containing $\IR^+\times G_{D+1},  \IR^+\times SO(d,d)$ and $\IR^+\times SL(d+1)$, respectively.
We analyze the Fourier coefficients of $\theta_{G_D}$ with respect to the unipotent radical
of $P$, and show that they can consistently be interpreted as 
\begin{itemize}
\item[i)] 
1/2-BPS particles in dimension $D+1$, whose Euclidean wordline winds around the extra
circle, together with Taub-NUT instantons when $D=3$; 
\item[ii)] Euclidean D-branes wrapping even-dimensional tori inside $T^{d}$, together with
Euclidean  NS5-branes on $T^6\subset T^7$ when $d\geq 7$ and Euclidean KK5-branes when $d=7$; 
\item[iii)]
Euclidean M2-branes on $T^3\subset T^{d+1}$ and Euclidean M5-branes on $T^6\subset T^{d+1}$,
together with KK6-branes when $D=3$.
\end{itemize}
Point i) realizes the general correspondence between BPS instantons in dimension $D$ and
black holes in dimension $D+1$ advocated in \cite{Gunaydin:2005mx,Alexandrov:2008gh},
by analogy with field theory \cite{Polyakov:1976fu,Seiberg:1996nz,Seiberg:1996ns}. 
We concentrate on the functional form of the Fourier coefficient (i.e. the Whittaker
and generalized Whittaker
vectors), leaving an analysis of the summation measure to future work. 

Before discussing each case in turn, it is useful to recall the general form of the particle 
multiplet in M-theory compactified on $T^{d}$ \cite{Obers:1998fb}\footnote{In the rest of this
section, indices $I,J,\dots$ run over $1,d$ in case i), or $1,\dots d+1$ in case iii);
indices $i,j,\dots$ run over $1,d-1$ in case i), or $1,\dots d$ in case ii).}. Point-like
particles in dimension $D=11-d\geq 4$,  arise as Kaluza-Klein (KK) particles, with momentum 
$m_I \in H^1(T^d,\IZ)$, M2-branes wrapped on a torus $T^2$ inside $T^d$, with charge $m^{IJ} \in H_2$, M5-branes wrapped on a torus $T^5$ inside $T^d$, with charge $m^{IJKLM} \in H_5$, 
and KK6-monopoles wrapped on a torus $T^6$, with charge $m^{I;7} 
\in  H_1\otimes H_7$ (for $d=8$,
there are additional exotic states that we do not need to consider here). These charges furnish
a finite-dimensional representation of the U-duality group $E_D$, as displayed in Table \ref{table1}.
BPS particles can preserve 1/2 or 1/4 fraction of the supersymmetry in  $D\geq 6$, or
1/2,1/4 or 1/8 fraction of the supersymmetry in  $D=4,5$ \cite{Ferrara:1997ci}. 
Requiring that the particle is 1/2-BPS puts quadratic conditions  on these charges \cite{Ferrara:1997ci,Obers:1998fb}, which take the schematic form
\begin{subequations}
\bea
\label{c1}
k^1 &=& m_1\, m^{2} = 0\ ,\qquad \\
\label{c2}
k^{4} &=&  m_1 \, m^{5} + m^2 m^2 =0\ ,\qquad \\ 
\label{c3}
k^{1;6} &=& m_1 m^{1;7} + m_2 m^5 = 0\ ,\qquad  \\
\label{c4}
k^{3;7} &=& m^2 m^{1;7} + m^5 m^5 = 0\ ,\qquad \\ 
\label{c5}
k^{6;7} &=& m^5 m^{1,7}\ .
\eea
\end{subequations}
These conditions themselves furnish
a finite-dimensional representation of $G_D$ which turns out to be the same as the 
string multiplet, displayed in Table \ref{table1}. In dimension $D\geq 6$, 1/4-BPS 
states can have arbitrary charges, while in dimension $D=4,5$, requiring that the
state is 1/4-BPS puts cubic conditions on the charges (the vanishing of the cubic
invariant of the $\irrep{27}$ representation of $E_{6(6)}$ for $D=5$, or the vanishing of the
differential of the quartic invariant of the  $\irrep{56}$ representation of $E_{7(7)}$
for $D=4$). In dimension $D=5$ (respectively $D=4$), 1/8-BPS states can have 
arbitrary charges, provided the cubic (respectively, quartic) invariant is positive.

It is also useful to recall that instanton effects are generally in correspondence with positive
roots of $G_D$ \cite{Obers:1998fb}.  In particular, in the ``M-theory limit" where the volume 
of  $T^{d+1}$ becomes large, they can be labelled by charges $m^{IJK}\oplus m^{IJKLMN} \oplus
m^{I;8}$ valued in $H_3 \oplus H_6 \oplus H_{1}\otimes H_8$, corresponding to Euclidean 
M2-branes, M5-branes and Kaluza-Klein monopoles wrapped on tori of the appropriate
dimension inside $T^{d+1}$.  In the ``string theory limit" where the $D$-dimensional 
string coupling $g_D = g_s / \sqrt{V_d/l_s^d}$ goes to zero, they instead decompose 
as D-instantons, with charges  transforming as a spinor $\Psi$
of $SO(d,d)$ (i.e.  an even polyform $m\oplus m^{ij}\oplus \dots$ of $SL(d)$ in type IIB), 
and NS5-brane
and KK5-brane instantons with charges transforming as $m^{ijklmn} \oplus m^{i;7}$
in $H_6\oplus H_1\otimes H_7$ (i.e. a singlet in $d=6$, or a vector of $SO(d,d)$ in $d=7$).
In the absence of NS-instantons, the 1/2-BPS condition on D-instantons is that their
charge should be a pure spinor of $SO(d,d)$ in Cartan's sense. In particular,
the number of independent charges describing 1/2-BPS instantons is 
$(d^2-d+2)/2$.

We now discuss the cases $D=6,5,4,3$ in turn, and analyze the non-perturbative effects
predicted by the conjecture \eqref{conj}.

\subsection{M-theory on $T^5$}

We start with $D=6$, where the U-duality group is $G_6=SO(5,5)$.

\subsubsection*{Decompactification limit}

The limit where one circle $S^1$ in $T^5$ becomes large is controlled by 
the branching of $SO(5,5)$ under
its maximal subgroup $\IR^+ \times SL(5)$, where the first factor corresponds to 
the non-compact Cartan generator $H_{\alpha_4}$ in the notations of  \cite{Kazhdan:2001nx}
and the second is identified as the U-duality group $G_7$ in 7 dimensions.
Under the action of this factor, the Lie algebra of $SO(5,5)$ decomposes as
\be
\label{3ge6d}
\irrep{45} = \irrep{10'}\vert_{-1} \oplus ( \irrep{24}+\irrep{1} )\vert_{0} \oplus  \irrep{10} \vert_{1} \ .
\ee
Accordingly, the moduli space $K_6\backslash G_6$ decomposes as 
$\IR^+ \times K_7\backslash G_7 \times \IR^{10}$, corresponding to the radius $R$
of the circle, the moduli in $D=7$,  parametrized by a (unit determinant,
symmetric matrix) coset representative $G^{AB}$ ($A,B=1\dots 5$) and the 
Wilson lines of the 10 gauge fields in 7 dimensions on the circle, denoted by 
the antisymmetric matrix $\Theta^{AB}$. In terms of representations 
of the modular group $SL(4)$ of $T^4$,
\be
G_{AB} = \frac{1}{l_M^6} \begin{pmatrix} (V^2/ l_M^6 )\, G^{IJ} + C^I C^J & C^I \\
C^J & 1 \end{pmatrix} 
\ee
where $G^{IJ}$ is the inverse metric on $T^4$ and $C^I=\epsilon^{IJKL} C_{JKL}$, 
and  $\Theta^{AB}$  consists of the mixed components
$G^{1I}$ and $C_{1IJ}$.

A representation of $\theta_{G_6}$ suited to 
this decomposition was given in 
\cite{Kazhdan:2001nx} , Eq. 4.6 \footnote{The $R$ dependence
in this formula can be found from the fact that the generator $\IR^+$ commuting with
$SL(5)$ is $-\frac{1}{4}(4H_0+2H_1+6H_2+5 H_3+3 H_4)=
y\pa_y+\sum_{i\in\{0,1,2,5\}} x_i \pa_{x_i}+\frac52$, in the notations of     \cite{Kazhdan:2001nx}.}
\be
\label{thd5}
\theta_{D_{5(5)}}\left( R, G^{AB}, \Theta^{AB} \right) = 
\sum_{m_{AB}} \mu(m_{AB})\, 
\frac{R^{3/2}\, e^{-2\pi R \sqrt{m_{AB}G^{AC}G^{BD} m_{CD} }
+2\pi \I  m_{AB} \Theta^{AB}}}{\sqrt{m_{AB}G^{AC}G^{BD} m_{CD} }} + \dots
 \ee
where $\mu(m_{AB})$ is given by the divisor sum 
 \be
 \label{mud5}
 \mu(m_{AB})=\sum_{d|m_{AB}} d
 \ee
and $m_{AB}$ is constrained to satisfy the rank 2 condition
\be
\label{crk2}
\epsilon_{ABCDE} \, m^{BC} m^{DE}=0\ .
\ee
The dots stand for ``degenerate" contributions independent 
of $\Theta^{AB}$, which should reduce to $R f_{R^4}^{(7)}$ plus some possible
power-like terms in $R$ (this decompactification limit is further analyzed in
Appendix A). The representation \eqref{thd5} corresponds to the Fourier 
decomposition of $\theta_{D_{5(5)}}$ with respect to the action of the Abelian group
$\irrep{10}\vert_1$ in \eqref{3ge6d}. 
With the above moduli identications, the integer matrix $m_{AB}$ can be interpreted
as labelling charges in the particle multiplet $\irrep{10}$  in dimension $D=7$.
The condition \eqref{crk2} is equivalent to the 1/2-BPS condition $k_1 = k^4=0$ 
where $k_1,k^4$ are the quadratic combinations of charges in \eqref{c1}, \eqref{c2} above.
For such a 1/2-BPS state, the  squared mass is given by $m_{AB}G^{AC}G^{BD} m_{CD}$
and its axionic couplings to the  off-diagonal metric
$G^{1I}$  and 3-form $C_{1IJ}$ are given by $m_{AB} \Theta^{AB}$. 
Thus, \eqref{thd5}
can be interpreted as a sum over 1/2-BPS instantons in $D=6$, which originate from 
1/2-BPS particles in $D=7$ whose wordline winds around the circle. 

\subsubsection*{String theory limit}

The limit of weak string coupling is instead controlled by the branching 
$SO(5,5) \supset \IR^+\times SO(4,4)$, where the first factor corresponds 
to the Cartan generator $H_{\alpha_1}$ in the notations of \cite{Kazhdan:2001nx}, and the second is  identified as the T-duality group on $T^4$. The Lie algebra of $SO(5,5) $
decomposes with respect to the $\IR^+$ action as a 3-grading,
\be
\label{3g45t}
\irrep{45} = \irrep{8'}\vert_{-1} \oplus ( \irrep{28}+\irrep{1} )\vert_{0} \oplus  \irrep{8} \vert_{1} \ .
\ee
Similarly, the moduli space $K_6\backslash G_6$ decomposes as 
$\IR^+\times SO(4)\times SO(4)\backslash SO(4,4)\times \IR^8$,
corresponding to the 6D dilaton $g_6$, Narain moduli and RR potentials $\Theta$ on $T^4$,
respectively. The latter transform as a spinor representation of $SO(4,4)$ (an even
polyform of $SL(4)$, in type IIB).

The presentation of $\theta_{D_{5(5)}}$ suited for this decomposition was not given
in \cite{Kazhdan:2001nx}, but it can be obtained easily by Fourier transforming
Eq. 4.28 in this paper with respect to $x_1$.  In this way, we arrive at \footnote{The $g_6$ dependence
in this formula can be found from the fact that the generator $\IR^+$ commuting with
$SO(4,4)$ is $-\frac12(2H_0+2H_1+2H_2+5 H_3+3 H_4)=y\pa_y+
\sum_{i\in\{0,1,2,3,4,5\}} x_i \pa_{x_i}+3$, in the notations of     \cite{Kazhdan:2001nx}.}
\be
\label{the6t}
\theta_{D_{5(5)}} = \sum_{\Psi} \mu(\Psi)\, \frac{K_1\left( \frac{1}{g_6} \sqrt{\Psi^2}\right) }
{g_6^2\, \sqrt{\Psi^2}} e^{\I \Psi \Theta} + \dots
\ee
where the dots denote degenerate contributions independent of the RR potentials $\Theta$,
which should reproduce the known tree-level and one-loop 
corrections\footnote{This has been demonstrated recently in \cite{Green:2010wi}.}. 
Here, the sum
runs over integer valued spinors $\Psi=(y,x_0,x_2,x_3,x_4,x_5,p_1, p_2)$
(in the notations of \cite{Kazhdan:2001nx})
subject to the pure spinor condition  
\be
\label{cso44}
x_0 p_1 - x_2 x_3 + x_4 x_5-p_2 y=0\ ,
\ee
and 
$\Psi^2$ denotes their $SO(4,4)$ invariant square norm, namely
\be
\Psi^2= y^2 + x_0^2 +  x_2^2 + x_3^2 + x_4^2 + x_5^2 + p_1^2 + p_2^2 
\ee
at the origin of the Narain moduli space. The fact that the Fourier coefficients have support on null charges  should come as no surprise, since the dimension of the minimal representation is 7.
These contributions can be interpreted as  1/2-BPS D-instantons of charge $\Psi$ wrapping $T^4$. 
In particular, there is no room for $e^{-1/g_6^2}$ effects, which made an appearance in the naive, 
unconstrained Epstein zeta series of \cite{Pioline:1997pu}.

\subsubsection*{M-theory limit}

Finally, one could analyze the expansion of $\theta_{D_{5(5)}}$ in the M-theory limit,
where the volume torus $T^5$ becomes large. The modular group of $T^5$ is related
to $G_4=SL(5,\IZ)$ by an outer automorphism of $G_5$, so corresponds to a 3-grading
of the same type as \eqref{3ge6d}, where now $\irrep{10}$ is identified as the 3-form
field $C_{IJK}$ on $T^5$. The Fourier  expansion takes the same form as in 
\eqref{thd5}, where each term is now interpreted as a contribution from Euclidean
M2-branes wrapping $T^5$ and the r\^ole of the radius $R$ is played by $(V_5)^{3/5}$.

\vskip 7mm

The fact that the Fourier coefficients of the minimal theta series have support
on 1/2-BPS charges, together with the matching of infinitesimal characters, strongly supports the conjecture that $f_{R^4}^{(6)}\propto \theta_{D_{5(5)}}$. It would be interesting to compare the measure 
\eqref{mud5} to the indexed degeneracies of 7-dimensional 1/2-BPS states.
The measure in the polarization \eqref{the6t}
is unknown at present, but could in principle be obtained from
the measure \eqref{mud5} by Poisson resummation. 
On general grounds, one would
expect that this measure is given by a product over all primes $p$ of the $p$-adic
Whittaker vector, which should be obtained by replacing the Bessel function $x^{-s} K_{s}(x)$
by its $p$-adic analogue $\mathcal{K}_{p,s}=(1-p^s |x|_p^{-s})/(1-p^s) 
\gamma_p(x)$ \cite{MR2290764}.
This  suggests that $\mu(\Psi)$ should be proportional to $\sum d^2$
where $d$ runs over common divisors of the entries in $\Psi$, but it would be good to
confirm this guess.

\subsection{M-theory on $T^6$}

We now turn to $D=5$.

\subsubsection*{Decompactification limit}

The limit where one circle inside $T^6$ becomes very large is controlled by the branching
$E_{6(6)}\supset \IR^+\times SO(5,5)$, where the first factor is the Cartan generator $H_{\alpha_5}$
in \cite{Kazhdan:2001nx}  and the second factor is identified as $G_6$. Under the $\IR^+$ action,
the Lie algebra of $E_{6(6)}$ decomposes into the 3-grading 
\be
\label{3ge6}
\irrep{78} = \irrep{16'}\vert_{-1} \oplus ( \irrep{45}+\irrep{1} )\vert_{0} \oplus  \irrep{16} \vert_{1} \ .
\ee
Similarly, the moduli space $K_5\backslash G_5$ decomposes as 
$\IR^+ \times K_5\backslash G_5 \times \IR^{16}$, corresponding to the radius $R$
of the circle, the $D=6$ moduli and the Wilson lines $\Theta$ of the 16 gauge fields in $D=6$.
The latter transform as a spinor of $G_6$.

A presentation of the minimal theta series of $G_5=E_{6(6)}(\IZ)$ suited to this
decomposition was given in \cite{Kazhdan:2001nx} , Eq. 4.47, in terms of a sum
over $5\times 5$ antisymmetric matrices $X_{AB}$ and an extra integer $y$. 
It was later recognized in \cite{MR2290764} that  the 11 integers $X_{AB}$ and $y$ parametrize 
a pure spinor $\Psi$ of $G_5$, which decomposes as the  even polyform $y \oplus X \oplus (X\wedge X/y)$ under $SL(5)\subset G_5$. Thus, we can write \footnote{The $R$ dependence
in this formula can be found from the fact that the generator $\IR^+$ commuting with
$SO(5,5)$ is $-\frac{1}{3}(  3 H_0 + 2 H_1 + 4 H_2 + 6 H_3 + 5 H_4 + 4 H_5) =y\pa_y+
\sum_{i\in\{0,1,2,3,4,5,7\}} x_i \pa_{x_i}+4$, in the notations of     \cite{Kazhdan:2001nx}.}
\be
\label{the6}
\theta_{E_{6(6)}} = \sum_{\Psi} \mu(\Psi)\, \frac{R^3\, K_1\left( R \sqrt{\Psi^2}\right)}{ \sqrt{\Psi^2}}  \,
e^{\I \Psi \Theta}
+ \dots
\ee
where the dots stand for some $\Theta$ independent terms, which should reproduce
$R f_{R^4}^{(6)}$ plus power-like terms in $R$. Here, $\mu(\Psi)$ is a certain moduli independent summation measure, and $\Psi^2$ is the 
$SO(5,5)$ invariant norm of the pure spinor $\Psi$. The spinor $\Psi$ is naturally interpreted
as the multiplet of charges for BPS particles in $D=6$  \cite{Dijkgraaf:1996cv}, 
and the purity condition is equivalent to the 1/2-BPS conditions \cite{Ferrara:1997ci}. 
Thus, \eqref{the6}
can be consistently interpreted as a sum over 1/2-BPS instantons in $D=5$, which originate from 
1/2-BPS particles in $D=6$ whose wordline winds around the circle. 

\subsubsection*{String theory limit}

Similarly, one could investigate  the Fourier decomposition of $\theta_{E_{6(6)}}$ with respect 
to the T-duality group $SO(5,5)$. Since the T-duality group and $G_5$ are related by an 
automorphism of the Dynkin diagram, the Fourier expansion takes the same form as  
\eqref{the6}, where now each term is identified as the contribution as a 1/2-BPS 
D-brane instanton on $T^5$. 

\subsubsection*{M-theory limit}

Let us now consider the M-theory limit of large $T^6$, corresponding to 
the branching $E_{6(6)}\supset SL(6)\times SL(2)$, where the last factor 
corresponds to the generators $E_{\pm \omega}, H_{\omega}$ in the notations
of \cite{Kazhdan:2001nx}. We shall denote this  factor by
$SL(2)_S$.Under the action of $H_{\omega}$, the Lie algebra
of  $E_{6(6)}$ decomposes into the 5-grading
\be
\label{5ge6}
\irrep{78} = \irrep{1}\vert_{-2} \oplus
\irrep{20}'\vert_{-1} \oplus
(\irrep{35}\oplus \irrep{1})\vert_{0} \oplus
\irrep{20}\vert_{1} \oplus
\irrep{1}\vert_{2} \ ,
\ee
where the grade $\pm 1$ spaces transform as an antisymmetric 3-form of $SL(6)$. 
 In contrast to the previous  cases, the positive grade generators do not commute, 
 but rather form a Heisenberg  algebra. This type of 5-grading will appear repeatedly
 in later examples, and we shall frame the discussion in a way which generalizes easily.
The appropriate language for this discussion is that of Jordan algebras of degree 3, 
but we shall avoid to use it explicitly (see e.g. \cite{Pioline:2006ni} for a review). 

In accordance with \eqref{5ge6}, the moduli space $K_5\backslash G_5$
decomposes as 
\be
USp(8)\backslash E_{6(6)} = \IR^{+} \times SO(6) \backslash SL(6) \times 
\IR^{20} \times \IR
\ee
corresponding to the $T^6$ volume, unit-volume metric, C-field $C_{IJK}$ on $T^6$
and finally, the six-form $E_{IJKLMN}$ dual to the C-field on $T^6$. It will be useful
to further decompose $SL(6)$ into $SL(3)\times SL(3)$, by grouping the 6 coordinates
on $T^6$ as $\{1,2,3\}\cup\{4,5,6\}$. In this way, the three-form decomposes as 
$\irrep{1}+\irrep{(3,3)}+\irrep{(3,3)}+\irrep{1}$, which we shall denote by 
$c^0, c^a, c_a, c_0$, where $a$ runs over pairs of indices $AA'$, where $A\in \{1,2,3\}$
and $A'\in \{4,5,6\}$. The commutant of $SL(3)\times SL(3)$ inside $G_5$ is $SL(3)$.
The latter is generated by $SL(2)_S$, the commutant of $SL(6)$, and by  another $SL(2)$
factor, corresponding to the generators $E_{\pm\beta_0},H_0$ in  \cite{Kazhdan:2001nx},
which we shall denote by $SL(2)_\tau$. The modular parameters associated to 
these two (non-commuting) $SL(2)$ actions are\footnote{These identifications are valid at linear 
order in the axionic couplings, and up to numerical coefficients. Accurate, convention-dependent formulae can be found in \cite{Pioline:2009qt}.}
\be
S = E_{123456}  + \I V_6/l_M^6\ ,\qquad \tau=C_{456} + \I V_{456}/l_M^3\ ,
\ee
where $V_{456}$ is the volume of the $T^3$ torus in the directions 456, denoted $T^3_{456}$.  
With a view to later generalizations, we shall denote $S=\sigma+\I e^\phi$, $\tau=\tau_1+\I\tau_2$.
We also allow for a non-vanishing value of $c_0=C_{123}$.
We shall restrict ourselves to the subset of the moduli space where only
these moduli are turned on (it is straightforward to generalize the formulae
to below at any point of $K_5\backslash G_5$). In practice, we are reduced to the
situation studied in \cite{Pioline:2009qt}, which the reader should consult for further
details.

Returning to the 5-graded decomposition \eqref{5ge6}, 
any automorphic form on $K_5\backslash G_5$ $/G_5(\IZ)$
can be decomposed into a sum of  Abelian and non-Abelian Fourier coefficients, corresponding to  representations of the Heisenberg group $\irrep{20}\vert_{1} \oplus
\irrep{1}\vert_{2}$ with trivial and non-trivial center $\irrep{1}\vert_{2}$, respectively. In the
present context,  these two types of contributions correspond to 
instantons with zero or non-zero M5-brane instanton charge, as we shall see presently.

The minimal theta series of $G_5=E_{6(6)}(\IZ)$ was expressed in \cite{Kazhdan:2001nx} , Eq. 4.43
in a manifestly $SL(3)\times SL(3)$-symmetric form as a 
sum over 11 integers, consisting of two singlets $y,x^0$ and  $\irrep{(3,3)}$
of $SL(3)\times SL(3)$,  denoted by $x^{a}$.  This sum corresponds to the non-Abelian 
Fourier coefficients with non-trivial
center $y$. In \cite{MR2094111,Pioline:2009qt} it was shown how to obtain the Abelian 
Fourier coefficients in the limit $y\to 0$. Putting these results together, we can write the 
minimal theta series of $E_{6(6)}$ as 
\be
\label{the6m}
\begin{split}
\theta_{E_{6(6)}} &= \sum_{x^0,x^{a}} \mu_A(x^0,x^{a})\, \frac{\tau_2}{(x^0)^2\, S_{0,x^0,x^{a}}^2} K_{1/2}(S_{0,x^0,x^{a}}) \\ 
&+ 
\sum_{y\neq 0} \sum_{x^0,x^{a}}  \mu_{NA}(y,x^0,x^{a}) \frac{\tau_2}{|x^0-\tau y|^2\, S_{y,x^0,x^{ij}}^2} 
K_{1/2}(S_{y,x^0,x^{a}}) e^{\I (y \sigma+x^0 c_0) -  \I \frac{(x^0 - y \tau_1) x }{y |x_0-y \tau|^2}} + \dots
\end{split}
\ee
where
\be
\label{sya}
S_{y,x^0,x^{a}} = \frac{e^{\phi}}{\sqrt{\tau_2}} \sqrt{ \frac{|x^0-\tau y|^2}{\tau_2} + (x^{ij})^2 + 
\frac{\tau_2 } {|x^0-\tau y|^2}\, (x_{a})^2+ \frac{(\tau_2)^2 } {|x^0-\tau y|^4} (x)^2}\ ,
\ee
and $x_{a}$ and $x$ are expressed in terms of $x^{a}$ via
\be
\label{ca0e6}
x_{a} = \frac12\kappa_{abc} x^{a} x^{b} \ ,\qquad 
x = \frac16 \kappa_{abc} x^a x^b x^c\ .
\ee
Here, $\kappa_{abc}$ is the invariant tensor in $\vee^3 \irrep{(3,3)}$ given by
$\kappa_{abc}=\epsilon_{ABC} \epsilon_{A'B'C'}.$

The  first and second lines 
in \eqref{the6m} correspond to the Abelian and non-Abelian Fourier coefficients,
respectively. For the Abelian part $y=0$, \eqref{sya} reduces to the action of 
Euclidean M2-branes wrapped on a three-cycle $m^{IJK}$ on $T^6$, 
which satisfies the 1/2-BPS condition $\irrep{20}\vee \irrep{20}\vert_{\irrep{35}}=0$.
Under the  decomposition $SL(6)\supset \IR^+ \times SL(3)\times SL(3)$,
$m^{IJK}$ decomposes as $x^0 \oplus x^{a} \oplus x_a/(x^0) \oplus x/(x^0)^2$
and the constraint is equivalent to \eqref{ca0e6}. 
Setting $x^0=1,x^{a}=0$, the action reduces to $e^\phi /\tau_2+\I c_0=
V_{123}/l_M^3+\I C_{123}$, which is the correct action for an M2-brane wrapped on $T^3_{123}$.

The non-Abelian contributions with $y\neq 0$ follow from the Abelian ones by an
action of $SL(2)_\tau$. Setting $y=1,x^0=0,x^{a}=0$, \eqref{sya} reduces to
$e^\phi+\I \sigma$, which is now recognized as the action of an M5-brane wrapped on $T^6$.
When all charges are switched on, there is an additional contribution to the phase 
proportional to $(x^0 - y \tau_1)/(y |x_0-y \tau|^2)$. The physical origin of this
coupling was explained in \cite{Pioline:2009qt}: in short, it is just the $SL(2,\IZ)$
image of the familiar axionic couplings, after absorbing a moduli independent (but charge
dependent) contribution into the measure $\mu_{NA}(y,x^0,x^a)$. Reinserting this phase
also allows to take the limit $y\to 0$ in the non-Abelian Fourier coefficients and recover the Abelian ones \cite{MR2094111,Pioline:2009qt}. It should be noted that the modified Bessel function
appearing in \eqref{the6m} has an exact semi-classical approximation
$K_{1/2}(x)e^{-x}\sqrt{2\pi/x}$, which indicates that quantum corrections around the
instanton background vanish beyond one-loop.

\vskip 7mm

The fact that the Fourier coefficients of the minimal theta series have support
on 1/2-BPS charges, together with the matching of infinitesimal character, 
again strongly supports the conjecture that $f_{R^4}^{(5)}\propto \theta_{E_{6(6)}}$.
If this conjecture indeed holds true, it would be very interesting to compute the
measure $\mu(y,X)$ and compare it to the indexed degeneracies of $1/2$-BPS
states in 6 dimensions \cite{Dijkgraaf:1996cv}. As before, one may guess from $p$-adic arguments
that $\mu(\Psi)$ should be proportional to $\sum d^2$ where $d$ runs over common 
divisors of the entries in $\Psi$, but it would be good to confirm this prediction.
The $p$-adic components of the measure $ \mu_{NA}(y,x^0,x^{a})$ 
were computed in \cite{MR2094111} for the polarization of interest, and it would be 
interesting to further elucidate it.

\subsection{M-theory on $T^7$}

We now turn to $D=4$.  

\subsubsection*{Decompactification limit}

The limit where the torus $T^7$ becomes large is controlled by the branching 
$E_{7(7)}\supset \IR^+ \times E_{6(6)} $, where the first factor corresponds to 
$H_{\alpha_6}$ in \cite{Kazhdan:2001nx}. Under the $\IR^+$ action, the Lie algebra of 
$G_4=E_{7(7)}$ decomposes as 
\be
\label{3ge7}
\irrep{133} = \irrep{27'}\vert_{-1} \oplus ( \irrep{78}+\irrep{1} )\vert_{0} \oplus  \irrep{27} \vert_{1} \ .
\ee
The moduli space $K_4\backslash G_4$ decomposes as $\IR^+ \times K_5\backslash G_5
\times \IR^{27}$, corresponding to the radius of the circle, 5D moduli and Wilson lines of the 
27 gauge fields $\Theta$ in $D=5$. The latter transform as a $\irrep{27}$ of $G_5=E_{6(6)}$.

A presentation of the  minimal theta series of $G_4=E_{7(7)}(\IZ)$ was written 
in \cite[Eq. 4.56]{Kazhdan:2001nx}\footnote{{\it Note added in v3:} 
The power of $J_4$ in the denominator of
Eq. 4.56 in \cite{Kazhdan:2001nx} should read $3/4$ rather than $5/4$, in agreement with \cite{DvorskySahi}. I am grateful to G. Bossard for pointing this misprint.}
  as a sum over a 16-dimensional spinor 
$\Psi=x_0 \oplus X \oplus *Y$ of $SO(5,5)$ (
where $X$ is a two-form and $*Y$ is the Hodge dual of a one-form) and an 
extra integer $y$. it was later recognized in \cite{MR2290764}  that the 17 
variables $(y,\Psi)$ transform as a $\irrep{27}$-dimensional representation $\Xi$ 
of $G_5=E_{6(6)}(\IR)$ satisfying the 
purity constraint $\irrep{27} \vee \irrep{27}\vert_{\irrep{27}}=0$. 
In terms of its decomposition under
$SO(5,5)$ and $SL(5)$, respectively,
\be
\irrep{27}=\irrep{1}+\irrep{16}+\irrep{10}=y \oplus (x^0\oplus X\oplus *Y)\oplus (x^0 Y/y \oplus
X \wedge X/y)\ .
\ee
In terms of this constrained variable, we may write\footnote{The 
$R$ dependence
in this formula can be found from the fact that the generator $\IR^+$ commuting with
$E_{6(6)}$ is $-\frac12(2 H_0 + 3 H_1 + 4 H_2 + 6 H_3 + 5 H_4 + 4 H_5 + 3 H_6)=y\pa_y+
\sum_{i\in\{0,1,2,3,4,5,6,7,8,11,14\}} x_i \pa_{x_i}+6$, in the notations of   \cite{Kazhdan:2001nx}.}
\be
\label{the7}
\theta_{E_{7(7)}} = \sum_{\Xi} \mu(\Xi)\, 
 \frac{R^{9/2}\, K_{3/2}\left( R \sqrt{\Xi^2}\right) }{[\Xi^2]^{3/4}} e^{\I \Xi\Theta}+ \dots
\ee
Here, $\Xi^2$ is the $G_5$-invariant squared norm of $\Xi$, $\mu(\Xi)$ is a moduli independent summation measure, presently unknown, and the dots stand for $\Theta$-independent contributions,
which should reduce to $R f_{R^4}^{(5)}$ plus power-like terms.
This representation provides the Fourier decomposition of $\theta_{E_{7(7)}}$ with respect to the 27-dimensional Abelian group corresponding to the last factor in \eqref{3ge7}.

As before, the representation $\irrep{27}$ of $E_{6(6)}$ can be interpreted as the 
particule multiplet in $D=5$, which consists of the 6 KK charges
$m_I$, 15 M2-brane charges $m^{IJ}$, and 6 M5-brane charges $m^{IJKLM}$.
The purity constraint $\irrep{27} \vee \irrep{27}=0$ is precisely the 1/2-BPS condition for 
BPS states in $D=5$ \cite{Ferrara:1997ci}.

\subsubsection*{String theory limit}

Let us now consider the weak coupling limit in string theory, corresponding to 
the branching $E_{7(7)}\supset SO(6,6)\times SL(2)_S$, where the last factor 
correspond to $E_{\pm\omega},H_{\omega}$ in the notations of  \cite{Kazhdan:2001nx}.
Under the action of $H_\omega$ the Lie algebra of $E_{7(7)}$ decomposes into the 5-grading
\be
\label{5ge7}
\irrep{133} = \irrep{1}\vert_{-2} \oplus
\irrep{32}'\vert_{-1} \oplus
(\irrep{66}\oplus \irrep{1})\vert_{0} \oplus
\irrep{32}\vert_{1} \oplus
\irrep{1}\vert_{2} \ ,
\ee
similar to \eqref{5ge6}. 
Accordingly, the moduli space $K_4\backslash G_4$
decomposes as 
\be
SU(8)\backslash E_{7(7)} = \IR^{+} \times [SO(6)\times SO(6)] \backslash SO(6,6) \times 
\IR^{32} \times \IR
\ee
corresponding to the 4D string coupling, Narain moduli, RR fields on $T^6$
and the NS-axion. It will be convenient to further break $SO(6,6)$ into $SL(6)$,
and decompose the grade 1 space as $\irrep{32}=1\oplus 15\oplus 15'\oplus 1$. The commutant
of $SL(6)$ inside $E_{7(7)}$ is generated by $SL(2)_S$ and $SL(2)_\tau$.
The corresponding complex modular parameters can be identified as 
the 4D and 10 complexified couplings \cite{Pioline:2009qt},
\be
S = \psi + \I/g_4^2\ ,\qquad \tau=c^0 + \I/g_s\ ,
\ee
where $c^0$ and $c_0$ are the RR 0-form and 6-form on $T^6$ and $\psi$ is the NS-axion.  
We continue to denote $S=\sigma+\I e^\phi$, $\tau=\tau_1+\I\tau_2$.
As before, $\theta_{E_{7(7)}}$ can be decomposed into a sum of  Abelian and non-Abelian Fourier coefficients, corresponding to  representations of the Heisenberg group with trivial and non-trivial center, respectively. As
we shall see presently, these two types of contributions now correspond to 
instantons with zero or non-zero NS5-brane charge, respectively.

The minimal theta series of $G_4=E_{7(7)}(\IZ)$ was expressed in \cite{Kazhdan:2001nx} , Eq. 4.53
in a manifestly $SL(6)$-symmetric form as a 
sum over 17 integers, consisting of two singlets $y,x^0$ and an antisymmetric two-form $\irrep{15}$
of $SL(6)$,  denoted by $x^{a}=x^{ij}$.  Following the same steps as in \eqref{the6m},  we can write 
\be
\label{the7t}
\begin{split}
\theta_{E_{7(7)}} &= \sum_{x^0,x^{a}} \mu_A(x^0,x^{a})\, \frac{\tau_2^{3/2}}{(x^0)^3\, S_{0,x^0,x^{a}}^2} K_1(S_{0,x^0,x^{a}}) \\ 
&+ 
\sum_{y\neq 0} \sum_{x^0,x^{a}}  \mu_{NA}(y,x^0,x^{a}) \frac{\tau_2^{3/2}}{|x^0-\tau y|^3\, S_{y,x^0,x^{a}}^2} 
K_1(S_{y,x^0,x^{a}}) e^{\I (y \sigma+x^0 c_0) -  \I \frac{(x^0 - y \tau_1) x }{y |x_0-y \tau|^2}} + \dots
\end{split}
\ee
where $S_{y,x^0,x^{a}}$ and $x_a, x$ are still given by \eqref{sya}, \eqref{ca0e6}.
The symmetric tensor $\kappa_{abc}$ appearing in \eqref{ca0e6} is now given
in $SL(6)$ invariant terms by $\kappa_{abc} = \epsilon_{ijklmn}$, when $a=(ij),b=(kl),c=(mn)$.

The  first and second lines 
in \eqref{the6m} correspond to the Abelian and non-Abelian Fourier coefficients,
respectively. For the Abelian part $y=0$, $S_{y,x^0,x^{a}} $ reduces to the action of 
Euclidean D-branes wrapped with charges $\Psi\in\irrep{32}$ on an  even cycle on $T^6$, 
which satisfies the 1/2-BPS condition $\irrep{32}\vee \irrep{32}\vert_{\irrep{66}}=0$.
Under the  decomposition $SO(6,6)\supset \IR^+ \times SL(6)$,
$\Psi$ decomposes as $x^0 \oplus x^{a} \oplus x_a/(x^0) \oplus x/(x^0)^2$
and the constraint is equivalent to \eqref{ca0e6}. 
Setting $x^0=1,x^{ij}=0$, the action reduces to $e^\phi /\tau_2+\I c_0$, 
which is the correct action for an D5-brane wrapped on $T^6$.
The non-Abelian contributions with $y\neq 0$ follow from the Abelian ones by an
action of 10D S-duality $SL(2)_\tau$. Setting $y=1,x^0=0,x^{a}=0$, $S_{y,x^0,x^{a}} $ reduces to
$1/g_4^2+\I \psi$, which is now recognized as the action of a NS5-brane wrapped on $T^6$.
The structure of these instanton corrections is essentially identical to the one found 
in \cite{Pioline:2009qt} in the study of quantum corrections to the hypermultiplet moduli space.

\subsubsection*{M-theory limit}

Finally, the M-theory limit corresponds to the 
5-grading 
\be
\label{5ge7m}
\irrep{133} = \irrep{7}\vert_{-2} \oplus
\irrep{35}\vert_{-1} \oplus
(\irrep{48}\oplus \irrep{1})\vert_{0} \oplus
\irrep{35}\vert_{1} \oplus
\irrep{7}\vert_{2} \ ,
\ee
which arises from the action of the first factor in the maximal subgroup 
$\IR^+ \times SL(7)$. The non-Abelian Fourier coefficients with respect 
to the action of the nilpotent group $\irrep{35}\vert_{1} \oplus
\irrep{7}\vert_{2}$ should encode the contributions of Euclidean M2 and M5 branes
wrapped on $T^3$ and $T^6$, but I have not attempted to analyze them.

\vskip 7mm

The fact that the Fourier coefficients of the minimal theta series have support
on 1/2-BPS charges, together with the matching of infinitesimal characters, strongly supports the conjecture that $f_{R^4}^{(4)}\propto \theta_{E_{7(7)}}$. It would be interesting to compare the measure 
in \eqref{the7} to the indexed degeneracies of 5-dimensional 1/2-BPS states \cite{Maldacena:1999bp}.
A naive $p$-adic argument suggests that $\mu(\Xi)$ should be proportional to $\sum d^3$
where $d$ runs over all divisors of the vector $\Xi$, it would be interesting to confirm this
by Fourier transforming the result in \cite{MR2094111}.

\subsection{M-theory on $T^8$}

\subsubsection*{Double decompactification limit}

Finally, we turn to $D=3$. Under the decompactification 
to $D=4$, the Lie algebra of $G_3=E_{8(8)}$ decomposes into the 5-grading
\be
\label{5ge8}
\irrep{248} = \irrep{1}\vert_{-2} \oplus
\irrep{56}\vert_{-1} \oplus
(\irrep{133}\oplus \irrep{1})\vert_{0} \oplus
\irrep{56}\vert_{1} \oplus
\irrep{1}\vert_{2} \ ,
\ee
where the index denotes the weight under the non-compact Cartan generator of 
the first factor in the $SL(2)\times E_{7}$ maximal subgroup. As in \eqref{5ge6} and
\eqref{5ge7},  the positive grade generators  form a Heisenberg algebra. 
Correspondingly, the $D=3$ moduli space $K_3\backslash G_3$
decomposes as 
\be
SO(16)\backslash E_{8(8)} = \IR^{+} \times SU(8)\backslash E_{7(7)} \times \IR^{56} \times \IR
\ee
corresponding to the radius of the circle, the 4D moduli, the Wilson lines of the 
56 gauge fields in $D=4$ and finally the NUT scalar $K_{1;1IJKLMNP}$.
As before, it will be useful to further break $E_{7(7)}$ down to $E_{6(6)}$,
whose commutant in $E_{8(8)}$ is an $SL(3)$ subgroup  generated by
$SL(2)_S$ and $SL(2)_\tau$. In effect, this corresponds to decompactifying
two directions $S_1\times S_1'$ of radii $R_1$ and $R_2$, such that the U-duality group is broken to $G_5=E_{6(6)}$. 
The $SL(2)_S$ and $SL(2)_\tau$ symmetries are then interpreted 
as Ehlers symmetry and the $T^2$ modular group, with respective  
modular parameters 
\be
S = K_{1;12345678} + \I R_1^2 R_2 V_6 /l_M^9\ ,\qquad \tau= \frac{g_{12}}{g_{22}}+ \I
\frac{\sqrt{g_{11}g_{22}-g_{12}^2}}{g_{11}}\ ,
\ee
the remaining axion being $c_0=K_{2;12345678}$.
We continue to denote $S=\sigma+\I e^\phi$, $\tau=\tau_1+\I\tau_2$. Under
this decomposition, the grade 1 space decomposes as  $\irrep{56}=1\oplus \irrep{27}
\oplus \irrep{27}'\oplus 1$.

A presentation of the minimal theta series of $G_3=E_{8(8)}(\IZ)$ suited to this 
decomposition (in fact, the only currently available presentation, as far as I know)
was given in \cite{Kazhdan:2001nx} , Eq. 4.69 as a 
sum over 29 integers, consisting of two singlets $y,x^0$ and a $\irrep{27}$ of $G_5=E_{6(6)}$, 
denoted by $x^a$. Following the same steps as before, we may write
\be
\label{the8}
\begin{split}
\theta_{E_{8(8)}} &= \sum_{x^0,x^a} \mu_A(x^0,x^a)\, \frac{\tau_2^{5/2}}{(x^0)^5\, S_{0,x^0,x^a}^2} K_2(S_{0,x^0,x^a}) \\ 
&+ 
\sum_{y\neq 0} \sum_{x^0,x^a}  \mu_{NA}(y,x^0,x^a) \frac{\tau_2^{5/2}}{|x^0-\tau y|^5\, S_{y,x^0,x^a}^2} 
K_2(S_{y,x^0,x^a}) e^{\I (y \sigma+x^0c_0) -  \I \frac{(x^0 - y \tau_1) x }{y |x_0-y \tau|^2}} + \dots
\end{split}
\ee
where $S_{y,x^0,x^{a}}$ and $x_a, x$ are still given by \eqref{sya}, \eqref{ca0e6}.
The symmetric tensor $\kappa_{abc}$ appearing in \eqref{ca0e6} is now the
invariant tensor in the symmetric tensor product $\vee^3 \irrep{27}$.

The  first and second lines 
in \eqref{the6m} correspond to the Abelian and non-Abelian Fourier coefficients,
respectively. For the Abelian part $y=0$, $S_{y,x^0,x^{a}} $ is recognized as 
the mass of 1/2-BPS states in $D=4$. From the 5D point of view, those originate
as KK-monopoles localized on $S_1'$, with charge $x^0$, 5D BPS
strings wound around $S_1'$, with charge $x^a$, 5D black holes
localized on $S_1'$, with charge $x_a$, and finally Kaluza-Klein states on $S_1'$.
The constraint \eqref{ca0e6} ensures that these 5D particles are 1/2-BPS. 
Setting $x^0=1,x^{a}=0$, the action reduces to $e^\phi /\tau_2+\I c_0=
R_2^2 R_1 V_6/ l_M^9+\I  K_{2;12345678} $, which is the correct action for
KK-monopole localized on $S_2'$. 
The non-Abelian contributions with $y\neq 0$ follow from the Abelian ones by an
action of  $SL(2)_\tau$, the modular group of $T^2$. Setting $y=1,x^0=0,x^{a}=0$, $S_{y,x^0,x^{a}} $ reduces to $K_{1;12345678} + \I R_1^2 R_2 V_6 /l_M^9$, which is now recognized as the action of
Taub-NUT instanton localized on $S^1$.

\subsubsection*{String theory and M-theory limits}

The weak string coupling limit corresponds to the branching $E_{8(8)}\supset \IR^+ \times
SO(7,7)$. The action of  $\IR^+$ leads to a 5-grading
\be
\label{5ge82}
\irrep{248} = \irrep{14}\vert_{-2} \oplus
\irrep{64}\vert_{-1} \oplus
(\irrep{91}\oplus \irrep{1})\vert_{0} \oplus
\irrep{64}\vert_{1} \oplus
\irrep{14}\vert_{2} \ .
\ee
 Similarly, the M-theory 
limit of large $T^8$ corresponds to the 7-grading
\be
\label{7ge8}
\irrep{248} = \irrep{8}\vert_{-3} \oplus \irrep{28}\vert_{-2} \oplus
\irrep{56}\vert_{-1} \oplus
(\irrep{63}\oplus \irrep{1})\vert_{0} \oplus
\irrep{56}\vert_{1} \oplus
\irrep{28}\vert_{2} \oplus \irrep{8}\vert_{3} \ .
\ee
associated to the branching $E_{8(8)}\supset \IR^+ \times
SL(8)$. It would be interesting to investigate the non-Abelian Fourier decomposition with respect to
the action of the nilpotent groups  $\irrep{64}\vert_{1} \oplus
\irrep{14}\vert_{2}$ and $\irrep{56}\vert_{1} \oplus
\irrep{28}\vert_{2} \irrep{8}\vert_{3}$, which however are 
no longer of Heisenberg type. The latter should encode the contributions of KK-monopole
instantons bound to Euclidean M2 and M5-branes.

\vskip 7mm

At any rate, the conjecture that $f_{R^4}^{(3)}\propto \theta_{E_{8(8)}}$ is in perfect agreement with 
expected non-perturbative contributions, coming from  1/2-BPS particle in $D=4$
and their bound states with Taub-NUT instantons.

\section{Comments on 1/4-BPS couplings \label{sec_d4r4}}

In this final section, we extend our discussion to the case of 1/4-BPS couplings.
By the same argument based on counting fermionic zero modes, 
we expect that 12-derivative couplings in the low
energy effective action around vacua with 32 supersymmetries (such as $\nabla^4 R^4$) 
should receive 
instanton corrections from 1/4-BPS configurations only. These should arise from 1/4-BPS particles 
in $D+1$ dimensions, supplemented by Taub-NUT instantons for $D=3$. In dimension 
$D+1\geq 6$, generic BPS states are 1/4-BPS states, so this condition puts no restriction
on the possible charges. For $D+1=5$ or $D+1=4$, however, 
as reviewed at the beginning of Section \ref{sec_np}, the 1/4-BPS condition requires
that the charges satisfy a cubic constraint, $\irrep{27}^3=0$ or $\irrep{56}^3\vert_{\irrep{56}}=0$,
respectively. 
This constraint should be reflected in the structure of the Fourier coefficients 
of the automorphic form of interest. 

In the former case $D=4$, the space of solutions to 
$\irrep{27}^3=0$ has dimension 26. One should therefore expect 
that $f_{\nabla^4 R^4}$ should 
be associated to an automorphic representation of $E_{7(7)}$ of functional dimension 26.
Interestingly, such a representation has been constructed in \cite{MR1327538,MR1421947},
by considering the continuation of the quaternionic discrete series\footnote{It should be stressed that 
\cite{MR1327538,MR1421947} deal with the quaternionic real form of the complex groups 
$E_6,E_7,E_8$. Their construction presumably also yields unitary representations 
of the same dimension for the split real form, but the corresponding automorphic forms
may be quite different, as they must be invariant under different maximal compact subgroups.}
 $\pi_k$ at the 
value $k=10$ (the minimal series of $E_{7(7)}$  itself  arises as the continuation at $k=6$).
This representation is characterized by the fact that its Abelian Fourier coefficients
with respect to the 5-grading \eqref{5ge7}
have support on the 25-dimensional, 1/4-BPS locus
$\irrep{32}^3\vert_{32}=0$.  By dimension counting, it must be the case that its Fourier 
coefficients with respect to the 3-grading \eqref{3ge7}  have support  
on the 5D 1/4-BPS charge orbit $\irrep{27}^3=0$. Moreover, its infinitesimal character
is $\rho-\frac{k}{2} \lambda_{\irrep{133}}=[-4,1,1,1,1,1,1]$, the same as that of the Epstein 
zeta series in the string multiplet of order $5/2$. This is consistent with the known fact that
the $\nabla^4 R^4$ couplings in dimension $D\geq 7$ are given (in part) by Epstein
zeta series of order $5/2$ \cite{Green:1999pu,Basu:2007ru}.

Similarly, for $D=3$, the space of solutions to $\irrep{56}^3\vert_{\irrep{56}}=0$ has dimension
45 \cite{Ferrara:1997uz}. Adding in the NUT charge, one should therefore expect that  $f_{\nabla^4 R^4}$ should be associated to a representation of functional dimension 46.
Fortunately, such a representation has been constructed in \cite{MR1327538,MR1421947},
by considering the continuation of the quaternionic discrete series $\pi_k$ at the value $k=18$ (the minimal series of $E_{8}$  itself  arises as the continuation at $k=10$),
and its Abelian Fourier coefficients are known to have support on the 4D 1/4-BPS charge
orbit $\irrep{56}^3\vert_{\irrep{56}}=0$. Its infinitesimal character
is $\rho-\frac{k}{2} \lambda_{\irrep{248}}=[1,1,1,1,1,1,1,-8]$, which happens to be in the same
Weyl orbit as $[-4,1,1,1,1,1,1,1]$. Thus, this representation has the same infinitesimal character
as the Epstein zeta series in the string multiplet with order $5/2$, in agreement with higher 
dimensional expectations. $\nabla^4 R^4$ couplings in dimension $D\geq 5$ 
can be obtained by decompactification, as discussed in Appendix A.


To conclude, constraints on the allowed instanton charges, eigenvalues 
under the ring of invariant differential operators and U-duality 
tightly constrain the possible non-perturbative completions of BPS-saturated amplitudes. 
For $R^4$ couplings in toroidal compactifications of
M-theory, we believe that these constraints uniquely select the minimal theta series
of $G_D$, which is known rather explicitly \cite{Kazhdan:2001nx,MR2094111}. 
It would be very interesting to check that known perturbative contributions
are correctly reproduced, and to compare the summation measure against the
indexed degeneracies of BPS black holes in dimension $D+1$ 
derived e.g. in \cite{Dijkgraaf:1996cv,Maldacena:1999bp}. For 1/4-BPS couplings
in dimension $D$, we have identified a candidate representation, but we do not know
of any automorphic form attached to it, nor whether this form might be unique.
At any rate, it would be interesting to compute the Whittaker vector, and see if it 
correctly reproduces the expected mass formula for $1/4$-BPS states. 
One may also ask if the same type of arguments applies to 1/8-BPS saturated amplitudes
in $D=3,4,5$, such as $\nabla^6 R^4$. Unfortunately, such amplitudes are no longer eigenmodes
of the Laplacian \cite{Green:2005ba}, and seem to inhabit a world yet uncharted by mathematicians.

\acknowledgments

The conjecture that the minimal theta series of $E_{8}$ should control the $R^4$ couplings
in M-theory on $T^8$ arose in discussions with A. Neitzke back in January 2007, based on 
the observation that its Abelian Fourier coefficients have support on 1/2-BPS charges. 
I am indebted to him and M. G\"unaydin, E. Kiritsis, N. Obers, D.Persson, A. Waldron
for collaboration on this or closely related issues.  
I am also grateful to  M. A. A. van Leeuwen for providing the LiE software, which was instrumental
in computing the Weyl orbits of the infinitesimal characters, and to the theory group at ETHZ for hospitality and providing a stimulating atmosphere.

\appendix

\section{Decompactification limits}

In this appendix, we study the behavior of our conjectural exact results 
under decompactification from $D$ to $D+1$ 
dimensions. The behavior of $f_{R^4}^{(D)}$ and $f_{\nabla R^4}^{(D)}$
in this limit has been recently spelled out in \cite{Green:2010wi,Green:2010sp},
while the behavior of the proposed automorphic forms can be studied using
results available in the mathematical literature \cite{MR1159110,MR1469105}
for the constant terms of certain degenerate Eisenstein series under various
parabolic subgroups. 

\subsection{Constrained Epstein Zeta series and degenerate Eisenstein series}

In \cite{MR1469105}, the minimal automorphic representation for all 
simply laced groups is obtained as a residue of a degenerate Eisenstein 
series 
\be
\label{degeis}
E^*_P(g,f,s) = L(G,P,s)\, \sum_{\gamma\in G(\IQ)\backslash P(\IQ)} \, f(\gamma g,s) 
\ee
where $P$ is a certain maximal parabolic subgroup of $G$, $f$ is a vector in the induced representation $I_P(s)=\Ind_P^G \chi_{P,s}$, 
$\chi_{P,s}$ is a family of characters of $P$, and $L(G,P,s)$ is a normalizing
factor such that $E^*_P(g,f,s)$ has a finite number of poles and satisfies a 
simple functional equation. We take $f$ to be the unique 
(suitably normalized) spherical vector $f_K$ in $I(s)$, though the construction extends
to arbitrary K-finite vectors whose restriction to $K$ is independent 
of $s$  \cite{MR1469105}. The infinitesimal character computed from 
the list of $(P,\chi_P(s))$ listed in   \cite{MR1469105} matches the 
infinitesimal character $\rho-2s'\lambda_\cR$ of 
a constrained Epstein Zeta series $\eis{G}{\cR}{s'}$ at $s'=\kappa(s+\frac12)$,
where the values of $(\cR,\kappa)$ are given in  Table 2\footnote{To see this, 
note that $\chi_{P,s}=|t|^{2\kappa(s+\frac12)}$ where $t$ parametrizes the $\IR^+$
factor in the Levi subgroup $L$ of $P(G)$ related to the parameter $|a|$ defined below via
$t=|a|$ for $G=D_m$, $t=|a|^4$ for $G=E_6$, $t=|a|^3$ for $G=E_7$, $t=|a|^2$ for $G=E_8$.}.
Thus, the degenerate Eisenstein series
\be
\label{defeisstar}
\eisstar{G}{\cR}{s}(g) \equiv E^*_{P} \left(g,f_K,\frac{s}{\kappa} - \frac12  \right) 
\ee
satisfies the same invariant differential equations as the constrained Epstein Zeta series
\eqref{defceis}, but has a finite number of poles and  satisfies the functional equation
\be
\label{degeisfunc}
\eisstar{G}{\cR}{s}(g) = \eisstar{G}{\cR'}{\kappa-s}(g) \ ,
\ee
where $\cR'$ appears in the last column of the table above. We conjecture
that $\eisstar{G}{\cR}{s}$ is in fact equal to $\eis{G}{\cR}{s}$, up to an $s$-
dependent factor, in the region of the $s$-plane where both series are
absolutely convergent. In contrast to  \eqref{defceis}, the meromorphic continuation
of \eqref{defeisstar} is well understood, and so are its constant terms
under various parabolic subgroups. In this appendix, we shall reformulate our
conjectures in terms of residues of \eqref{defeisstar} for suitable choices of $\cR,s$.

\begin{table}
$$
\begin{array}{cclcccc}
\hline
G & P & {\rm Levi}(P) & \kappa & \cR & \cR' \\ 
\hline
SO(d,d) &   Q(D_d) & \IR^+ \times SO(d-1,d-1) & d-1 & \vect & \vect \\
               &   P(D_d) & \IR^+ \times SL(d) & d-1 & \spi & \spb \\
               &   P_\alpha(D_d) & \IR^+ \times SL(d) & d-1 & \spb &\spi \\
\hline
E_{6(6)}&   P(E_6) & \IR^+ \times SO(5,5) & 6 &\irrep{27'} & \irrep{27}\\
               &   Q(E_6) & \IR^+ \times SO(5,5)  & 6 &\irrep{27} & \irrep{27'} \\
\hline
E_{7(7)}&   P(E_7) & \IR^+ \times E_{6(6)} & 9  &\irrep{56}  & \irrep{56}\\
               &   P_{\rm Heis}(E_7) & \IR^+ \times SL(2) \times SO(6,6) 
& 17/2 &\irrep{133} & \irrep{133} \\
\hline
E_{8(8)}&   P(E_8) &  \IR^+ \times E_{7(7)}  & 29/2  &\irrep{248} &\irrep{249}  \\
\hline
\end{array}
$$
\caption{Dictionary between degenerate Eisenstein series and Epstein Zeta series}
\end{table}

\subsection{Constant terms}

The constant term of $\eis{E_{d}}{\cR}{s}$ with respect to the parabolic subgroup 
$Q(E_{d-1})$ was analyzed in  the course of the proof of Thm. 2.3 in \cite{MR1469105}.
Translating to our notations, this may be summarized as follows:\footnote{For the
$SO(d,d)$ case, we also rely on \cite{MR1159110,Green:2010wi}.}, 
\bse
\label{ctedegeis}
\be
\begin{split}
\int_{N(Q)} \eisstar{SO(d,d)}{\vect}{s} =&
 |a|^{2s}\, \zetastar(2s+2-d)\, \eisstar{SL(d)}{\bar d}{s} \\
 &+|a|^{2(d-1-s)}\, \zetastar(2s+1-d)\,  \eisstar{SL(d)}{d}{s+1-\frac{d}{2}} \ ,
 \end{split}
\ee 
\be
\begin{split}
\int_{N(Q)} \eisstar{SO(d,d)}{\spi}{s} =& |a|^{2s}\, \eisstar{SO(d-1,d-1)}{\spi}{s} \\
&+ |a|^{2(d-1-s)}\, \eisstar{SO(d-1,d-1)}{\spi}{s-1} 
\ \qquad ( d\, {\rm odd} )
\end{split}
\ee 
\be
\begin{split}
\int_{N(Q)} \eisstar{SO(d,d)}{\spi}{s} =&|a|^{2s}\, \zetastar(2s+2-d)\, \eisstar{SO(d-1,d-1)}{\spi}{s} \\
&+ |a|^{2(d-1-s)}\, \zetastar(2s+1-d)\, \eisstar{SO(d-1,d-1)}{\spi}{s-1} 
 \ \qquad ( d\, {\rm even} )
 \end{split}
\ee 
\be
\begin{split}
\int_{N(Q)} \eisstar{E_{6(6)}}{27}{s} =& 
|a|^{4s}\, \eisstar{SO(5,5)}{\vect}{s}+  |a|^{15-2s}\, \eisstar{SO(5,5)}{\spb}{s-\frac32} \\
&+ |a|^{8(6-s)}\, \zetastar(2s-8)\, \zetastar(2s-11)
\end{split}
\ee
\be
\begin{split}
\int_{N(Q)} \eisstar{E_{6(6)}}{27'}{s} =& |a|^{8s}\, \zetastar(2s)\, \zetastar(2s-3) 
+ |a|^{2s+3}\, \eisstar{SO(5,5)}{\spi}{s-\frac12} \\
 &+ |a|^{4(6-s)}\, \eisstar{SO(5,5)}{\vect}{s-2}
 \end{split}
 \ee
\be
\begin{split}
\int_{N(Q)} \eisstar{E_{7(7)}}{56}{s} =& 
 |a|^{6s}\, \zetastar(2s)\, \zetastar(2s-4)\, \zetastar(2s-8)\\
 &+|a|^{2(s+1)}\, \zetastar(2s-8)\,  \eisstar{E_{6(6)}}{27'}{s-\frac12}\\
&+|a|^{2(10-s)} \, \zetastar(2s-9)\,  \eisstar{E_{6(6)}}{27}{s-\frac52}\\
&+|a|^{6(9-s)} \zetastar(2s-9)\, \zetastar(2s-13) \, \zetastar(2s-17)   
\end{split}
\ee
\be
\begin{split}
\int_{N(Q)} \eisstar{E_{8(8)}}{248}{s} =&  
|a|^{4s}\, \zetastar(2s)\, \zetastar(2s-5)\, \zetastar(2s-9)\, \zetastar(4s-28)
\\
&+ |a|^{2s+1} \, \zetastar(4s-28)\,  \eisstar{E_{7(7)}}{56}{s-\frac12}\\
&+  |a|^{12}\,  \eisstar{E_{7(7)}}{133}{s-3}\\
&+|a|^{2(15-s)} \, \zetastar(4s-29)\,  \eisstar{E_{7(7)}}{56}{s-5}
\\
&+|a|^{2(29-2s)} \zetastar(2s-19)\, \zetastar(2s-23) \, \zetastar(2s-28) \, \zetastar(4s-29) 
\end{split}
\ee
\ese
where $|a|$ parametrizes the $\IR^+$ factor in the Levi subgroup $L$ of $P$. In 
these formulae, $\zetastar(s)\equiv \pi^{-s/2} \Gamma(s/2) \zeta(s)$ is the 
completed Riemann zeta function, which satisfies $\zetastar(s)=\zetastar(1-s)$, 
$\Res_{s=1}\zetastar(s)=1=-\Res_{s=0}\zetastar(s)$, and is analytic away from $s=0,1$.
Note in particular that in each cases, the functional relation is manifest,
as the various terms get permuted under $s\to \kappa-s$. 

Let us also record the relevant normalizing factors:
\bse
\bea
L(D_d,P(D_d),s) &=& \prod_{k=1}^{\lfloor d/2 \rfloor}\, 
\zetastar(2s+2-2k)\\
L(D_d,Q(D_d),s) &=& 
\zetastar(2s) \, \zetastar(2s+2-d) \\
L(E_6,P(E_6),s) &=& \zetastar(2s)\,  \zetastar(2s-3)\\
L(E_7,P(E_7),s) &=& \zetastar(2s)\,  \zetastar(2s-4) \, \zetastar(2s-8)\\
L(E_7,P_{\rm Heis}(E_7),s) &=& \zetastar(2s)\,  \zetastar(2s-3) \, \zetastar(2s-5)
\, \zetastar(4s-16)\\
L(E_8,P(E_8),s) &=& \zetastar(2s)\,  \zetastar(2s-5) \, \zetastar(2s-9)\, \zetastar(4s-28)
\eea
\ese

Using these relations recursively, and assuming that all poles show up in the constant terms, 
we conclude that 
\begin{itemize}

\item $\eisstar{SO(d,d)}{\vect}{s}$ has simple poles at $s=0,\frac{d}{2}-1, \frac{d}{2}, d-1$;
the minimal theta series arises as the residue at $s=\frac{d}{2}-1$ (or $s=\frac{d}{2}$).

\item $\eisstar{SO(d,d)}{\spi}{s}$ and $\eisstar{SO(d,d)}{\spb}{s}$
have simple poles at most at $s=0,1,2,\dots d-1$ (excluding the 
value $s=\frac{d-1}{2}$ for $d$ odd, moreover, for $d$ even it vanishes at that value);
the minimal theta series arises as the residue at $s=1$ (or $s=d-2$).

\item $\eisstar{E_{6(6)}}{27'}{s}$ has simple poles at most at 
$s=0,\frac12,\frac32, 2, \frac52,\frac72, 4, \frac92, \frac{11}{2},6$;
the minimal theta series arises as the residue at $s=\frac{3}{2}$ (or $s=\frac{9}{2}$). The residue
at the apparent pole $s=\frac{5}{2}$ (or $s=\frac{7}{2}$) vanishes, as we shall see later.

\item $\eisstar{E_{7(7)}}{56}{s}$ has simple poles at most at all
half integers between $0$ and $9$; the minimal theta series arises as the 
residue at $s=2$ (or $s=7$). The unipotent representation of functional dimension
26 mentioned in Sec. 5 arises as the residue at $s=4$ (or $s=5$).

\item $\eisstar{E_{8(8)}}{248}{s}$ has simple poles at most at all
half integers between $0$ and $\frac{29}{2}$ and at $s=\frac{29}{4}$. 
the minimal theta series arises as the 
residue at $s=\frac{5}{2}$ (or $s=12$). The unipotent representation of functional dimension
46 mentioned in Sec. 5 arises as the residue at $s=\frac{9}{2}$ (or $s=10$).

\end{itemize}

\subsection{Decompactification limits for $R^4$ couplings}

Using \eqref{ctedegeis}, 
we may extract the constant term under $Q(G)$ of the minimal theta series $\theta_G$,
obtained as the residue of the  degenerate Eisenstein series at the value of $(\cR,s)$ indicated
above:
\bse
\be
\int_{N(Q)} \ \Res_{s=\frac52} \eisstar{E_{8(8)}}{248}{s} 
= \zetastar(19) \, \left[ |a|^6 \, \Res_{s=\frac52} \eisstar{E_{7(7)}}{56}{s}  
-|a|^{10}  [ \zetastar(5) ]^2 \, \right]\ ,
\ee
\be
\int_{N(Q)} \ \Res_{s=2} \eisstar{E_{7(7)}}{56}{s} 
= \zetastar(5) \, \left[ |a|^6 \, \Res_{s=\frac32} \eisstar{E_{6(6)}}{27'}{s}  
- |a|^{12}\, \zetastar(4) \,  \right]\ ,
\ee
\be
\int_{N(Q)} \ \Res_{s=\frac32} \eisstar{E_{6(6)}}{27'}{s} 
= |a|^6 \, \Res_{s=1} \eisstar{SO(5,5)}{\spi}{s}  
- |a|^{12} \, \zetastar(3)  \ ,
\ee
\be
\int_{N(Q)} \ \Res_{s=\frac32} \eisstar{SO(5,5)}{\vect}{3/2} 
=- |a|^3 \, \eisstar{SL(5)}{\bar 5}{3/2}  
- |a|^{5} \, \zetastar(2) \ . 
\ee
\ese
In the last equation, we used the fact that $\Res_{s=0} \eisstar{SL(n)}{n}{s}=-1$.

Let us now compare with the decompactification limit of $R^4$ couplings \cite{Green:2010sp},
\be
\label{decr4}
\int_{N(Q)} f_{R^4}^{(D)} =  \left( \frac{R}{l_{D+1}} \right)^{\frac{8-D}{D-2}}\,  
\left[ \frac{R}{l_{D+1}} \,  f_{R^4}^{(D+1)} + 
a_D\,  \left( \frac{R}{l_{D+1}} \right)^{8-D} \right]\ ,
\ee
where the last term is required for reproducing the massless threshold in dimension $D+1$
(for $D=7,8$, it must be multiplied by $\log R$; the prefactor $(l_D/l_{D+1})^{D-8}$ arises
from the change of units from $D+1$ to $D$-dimensional Planck length). 
Identifying $R/l_{D+1}=|a|^{D-2}$ for $D=3,4,5$, $R/l_{D+1}=|a|^2$ for $D=6$, and,
up to overall normalization,
\be
\begin{split}
 f_{R^4}^{(7)} &=  - \eisstar{SL(5)}{\bar 5}{3/2}  \ ,\qquad
 f_{R^4}^{(6)} =  \Res_{s=\frac32} \eisstar{SO(5,5)}{\vect}{s}  \ ,\qquad
 f_{R^4}^{(5)} = \Res_{s=\frac32} \eisstar{E_{6(6)}}{27'}{s} \ ,\\
  f_{R^4}^{(4)} &= \frac{1}{\zetastar(5)} \Res_{s=2} \eisstar{E_{7(7)}}{56}{s} \ ,\qquad
   f_{R^4}^{(3)} = \frac{1}{\zetastar(5)\, \zetastar(19) } \Res_{s=\frac52} \eisstar{E_{8(8)}}{248}{s} \ ,
\end{split}
\ee
we see that \eqref{decr4} is obeyed, provided 
\be
\label{so55s1v32}
 \Res_{s=1} \eisstar{SO(5,5)}{\spi}{s}   
 \propto  \Res_{s=\frac32} 
 \eisstar{SO(5,5)}{\vect}{s}  \ .
\ee
This identity is consistent with the values of the infinitesimal characters, 
and could in principle be further checked by comparing the constant
terms under both $P(G)$ and $Q(G)$, although this information is
not available at this point. The relations above are consistent with the 
conjectures in \cite{Obers:1999um,Green:2010wi,Green:2010sp}. 
Moreover, the coefficient $a_D$ is proportional to $\zetastar(8-D)$, as expected.

\subsection{Decompactification limits for $\nabla^4 R^4$ couplings}

Let us now evaluate the constant term for the residue of the degenerate Eisenstein
series at the value of $(\cR,s)$ proposed in  Section 5 to describe 1/4-BPS $\nabla^4 R^4$ couplings,
\bse
\be
\begin{split}
\Res_{s=9/2} \int_{N(Q)} \eisstar{E_{8(8)}}{248}{s} &= 
 |a|^{10} \, \zetastar(11)\,  \Res_{s=4} \eisstar{E_{7(7)}}{56}{s}
 + |a|^{12}\,  \Res_{s=3/2}\, \eisstar{E_{7(7)}}{133}{3/2} \\
&- |a|^{18}\,   \zetastar(4)\, \zetastar(9)\,\zetastar(11) \, ,
\end{split}
\ee
\be
\begin{split}
\Res_{s=4} \int_{N(Q)} \eisstar{E_{7(7)}}{56}{s} &= 
 -|a|^{10}\,   \eisstar{E_{6(6)}}{27'}{7/2}\\&+ 
|a|^{12} \, \zetastar(2)\,  \Res_{s=3/2} \eisstar{E_{6(6)}}{27}{s} 
- |a|^{24}\, \zetastar(4)\, \zetastar(8)\, ,
\end{split}
\ee
\be
\begin{split}
\int_{N(Q)} \eisstar{E_{6(6)}}{27'}{7/2} = &
|a|^{10}\,  
\left( \eishatstar{SO(5,5)}{\vect}{3/2} +   \eishatstar{SO(5,5)}{\spb}{1} \right) \\
&-2 |a|^{10} \log |a| 
\Res_{s=\frac32}  \eisstar{SO(5,5)}{\vect}{s} 
+ |a|^{28}\, \zetastar(4)\, \zetastar(7)\ .
\end{split}
\ee
\ese
In this last equation, we have used the fact that due to \eqref{so55s1v32}
(more precisely its image under the outer automorphism which exchanges
$\spi$ and $\spb$), the residue of the apparent pole of $\eisstar{E_{6(6)}}{27'}{s}$
at $s=7/2$ actually vanishes.  Moreover, we  denoted by a hat the regularized series 
after subtracting the pole, 
\be
\eishat{SO(5,5)}{\vect}{s} = \frac{1}{s-\frac32}  
\left(  \Res_{s=\frac32}  \eisstar{SO(5,5)}{\vect}{s} \right)
+ \eishatstar{SO(5,5)}{\vect}{\frac32} + \cO( s- \frac32)\ ,\ {\rm etc.}
\ee

Let us now compare to the  decompactification limit of $\nabla^4 R^4$
 couplings \cite{Green:2010sp},
\be
\label{decompd4r4}
\int_{N(Q)} f_{\nabla^4 R^4}^{(D)} =  \left( \frac{R}{l_{D+1}} \right)^{\frac{12-D}{D-2}}
\left[ \frac{R}{l_{D+1}} \,  f_{\nabla^4 R^4}^{(D+1)} 
+ b_D \left( \frac{R}{l_{D+1}}\right)^{6-D} \,  f_{R^4}^{(D+1)}  
+ c_D \left( \frac{R}{l_{D+1}} \right)^{12-D} \right]\ ,
\ee
where the second term must be multiplied by  $\log R$ for $D=5$.
Identifying, up to overall normalization,
 \be
\begin{split}
 f_{\nabla^4  R^4}^{(6)} &=   \eishatstar{SO(5,5)}{\vect}{\frac32}
 +\eishat{SO(5,5)}{\spb}{1}   \ ,\quad
 f_{\nabla^4 R^4}^{(5)} =  \eisstar{E_{6(6)}}{27'}{\frac72} \ ,\\
  f_{\nabla^4 R^4}^{(4)} &=  \Res_{s=4} \eisstar{E_{7(7)}}{56}{s} \ ,\qquad
   f_{\nabla^4 R^4}^{(3)} = \frac{1}{\zetastar(11) } \Res_{s=\frac92} \eisstar{E_{8(8)}}{248}{s} \ ,
\end{split}
 \ee
 we find agreement with \eqref{decompd4r4} for $D\leq 5$, with 
 $b_D\propto \zetastar(6-D)$, $c_D \propto \zetastar(12-D)$, 
 provided the following relations hold true:
 \be
  \Res_{s=3/2} \eisstar{E_{6(6)}}{27}{s} \propto  \Res_{s=3/2} \eisstar{E_{6(6)}}{27'}{s} \ ,
 \ee
\be
\Res_{s=2}\, \eisstar{E_{7(7)}}{56}{s} \propto \, \Res_{s=3/2}\, \eisstar{E_{7(7)}}{133}{s} \ .
\ee
As  evidence for the first relation, note that  the infinitesimal
 characters match, and so do the constant terms under $Q(E_6)$ provided 
 \eqref{so55s1v32} is obeyed and $\Res_{s=0} \eisstar{SO(5,5)}{\spb}{s}=
- \zetastar(3)$. As for the second relation, it is known that the minimal automorphic theta 
series of $E_7$ arises as a submodule of $I_{P_{\rm Heis}}(s)$ 
for $s=3/2$ (\cite{MR1266113} and \cite{MR1469105}, .prop. 4.1).
In order to compute the proportionality constant, one would need
to know the constant term under $Q(E_7)$.
In order to check \eqref{decompd4r4} for $D=6$, we 
would also need to know the constant term of $\eishat{SO(5,5)}{\spb}{s}$ under $P(SO(5,5))$.
These constant term computations could in principle be done following the method in \cite{MR1159110,MR1469105}, though I have not attempted to do so.


\providecommand{\href}[2]{#2}\begingroup\raggedright\endgroup

\end{document}